# Advancing Operational Efficiency: Airspace Users' Perspective on Trajectory-Based Operations

Pablo Costas Álvarez
*Airspace Operational Efficiency*
Boeing Research & Technology
Madrid, Spain
pablo.costas@boeing.com

Ítalo Romani de Oliveira
*Airspace Operational Efficiency*
Boeing Research & Technology
São José dos Campos, Brazil
italo.romanideoliveira@boeing.com

*Abstract*— This work examines the evolution of the Flight Operations Center (FOC) and flight trajectory exchange tools within the framework of Trajectory-Based Operations (TBO), emphasizing the benefits of the ICAO's Flight and Flow Information for a Collaborative Environment (FF-ICE) messaging framework and the use of Electronic Flight Bags (EFBs). It highlights the collaborative generation and management of four-dimensional flight trajectories, which serve as a common reference for decision-making among stakeholders, including Air Navigation Service Providers (ANSPs), airspace users and airport operators. Key enabling technologies such as Performance Based Navigation (PBN), data communications, and System-wide Information Management (SWIM) are discussed, illustrating their roles in facilitating rapid information exchange and trajectory optimization. A live flight case study is presented, successfully demonstrating TBO concepts through international collaboration, and pointing to significant improvements in safety, efficiency, and sustainability.

The paper presents key technical results achieved through TBO prototype implementations, including enhanced trajectory accuracy, improved flight path efficiency, and the ability to accommodate real-time adjustments based on evolving conditions. The integration of advanced trajectory optimization engines and automation capabilities within the FOC has led to more effective flight planning and execution, enabling airlines to negotiate trajectory changes dynamically, adapting to changing circumstances and optimizing operations throughout the flight lifecycle.

The findings indicate that TBO can significantly enhance operational predictability, flexibility, and strategic planning while reducing uncertainty and improving alignment between strategic and tactical actions. The main conclusions are: that TBO is already possible with most of the currently flying commercial aircraft; the full implementation of TBO is a feasible way of achieving a greener and more efficient aviation industry, with the potential for widespread benefits across the global airspace system; and that continued collaboration among stakeholders is key for further development and fruition of TBO.

*Keywords—TBO, FF-ICE, SWIM, Connected Aircraft, EFB*

## I. Introduction

ICAO defines Trajectory Based Operations (TBO) [1] as an Air Traffic Management paradigm aiming to align the flown flight path closely with user preferences, reducing conflicts, and resolving demand/capacity imbalances efficiently. In such an environment, a four-dimensional flight trajectory, collaboratively developed, negotiated, managed and shared serves as a common reference for decision-making across all stakeholders; in short, coordination and collaboration occur using the trajectory as the common plan for the flight.

TBO's role in unlocking higher Operational Efficiency can be achieved at different phases thanks to the rapid exchange of information, from the planning stage with improved strategic planning for enhanced predictability, to the tactical phase where there is decreased uncertainty because of the alignment of strategic plans and tactical actions. This all contributes to minimizing airspace delays, disruptions and environmental impact.

Key TBO-enabling technologies on the aircraft, portable electronic devices, and airline systems include Performance Based Navigation (PBN) for enhancing trajectory accuracy; data communications and aircraft connectivity to facilitate enhanced data exchanges; and SWIM (System-wide Information Management) [2] for automating the sharing of common trajectory and flight information.

The focus of this paper is set on the Airspace User, targeting automation, enhanced flight efficiency and operator experience. In many cases, Airspace Users have invested in powerful tools to develop trajectories optimized to their best interests; TBO will help them to meet these interests. One of notable aspects is that TBO increases the flexibility to plan and execute more economical and sustainable routes, by offering more accurate and higher-resolution descriptions of the 4D trajectories, especially by precisely indicating the location of the altitude change points and their types [3]. By decreasing the ambiguity in these descriptions, the trajectories can be quickly evaluated by all stakeholders and quickly adjusted. Because of that, we will showcase how optimization trade-off studies can be performed in the TBO environment. First, we submit a same flight mission to optimization in different optimizers; second, we compare the respective optimized routes according to certain criteria; and third, we submit their resulting flight plans to FF-ICE validation and analyze which solution has higher acceptability, where FF-ICE stands for Flight and Flow Information for a Collaborative Environment [4]. And, although several of these steps are still performed manually, it will help to elicit requirements and system performance criteria for designing more efficient flight planning systems in the TBO environment.

Recently flight-tested through the Boeing ecoDemonstrator Explorer 2023 program [5], these tools showcased the immense potential of TBO, with successful international collaboration leading to significant improvements in safety and efficiency. The demonstrated benefits include minimized delays, disruptions, and notable reductions in travel costs, time, and fuel usage, aligning with Boeing's commitment to sustainable aviation and optimizing operational efficiency, which can reduce carbon emissions up to 10%.

The full implementation of TBO emerges as a cornerstone for a greener more efficient aviation industry, Initial TBO benefits have been recently presented by different ANSP's, and airlines in the Global TBO Symposium [6].

## II. THE ROLES OF THE FLIGHT OPERATIONS CENTRER (FOC) AND OF THE ELECTRONIC FLIGHT BAG (EFB) IN TBO

In terms of TBO, ICAO's vision is to achieve an interoperable global air traffic management system for all users during all phases of flight that meets agreed levels of safety, provides for optimum economic operations, is environmentally sustainable and meets national security requirements.

While working on harmonized goals, TBO implementation is based on getting incremental benefits in near-term by using building blocks already available in the industry.

In this scope, the key enablers for TBO are:

- SWIM (2005);
- Connected Aircraft (2014);
- CPDLC Clearance (2003); and
- FF-ICE (2012).

These years correspond to the official release of their Operational Concept documents. These elements together contribute to synchronization of trajectories in a way that the same reference is available for all stakeholders during the flight planning and execution stages. This is a key feature of TBO, which is implemented following FF-ICE data standard, exchanged over the SWIM services architecture.

The framework depicted in Figure 1 was initially introduced in ATMRPP WP-793 to outline the essential processes that would be further developed for FF-ICE. Before departure, the processes established for FF-ICE/Release 1 are utilized to create an Agreed Trajectory, which serves as the foundation for the filed flight plan. Throughout the flight, the flight is provided with clearances that align with the Agreed Trajectory. However, as circumstances evolve (such as tactical interventions, changes in wind/temperature, or decisions made by the Airspace User), the Agreed Trajectory must be managed. This may involve negotiating changes following FF-ICE/Release 2 and obtaining clearances for some of these modifications.

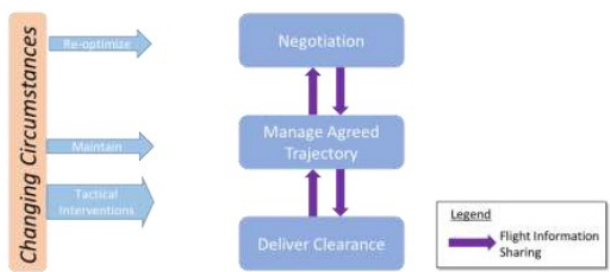

Figure 1: FF-ICE key processes

As defined in ICAO Manual on Flight and Flow for a Collaborative environment and refined by the ATMRPP-WG39, the Airspace User (eAU) involved in trajectory management and negotiation can be located in the Airline Flight Operations Center (FOC) but also in the flight deck. At the same time these actions can be supported by dispatchers (in the FOC) and flight crew (in the flight deck) but also assisted by automation, another key element in TBO.

The EFB supports trajectory negotiation and management using FF-ICE/R2 and has the ability to interface the Flight Management System by using standard Aircraft Interface Device (AID) in order to get trajectories from avionics [7]. This process is not the only way for ground systems to get trajectories from the aircraft, as one can use Datalink communications (ground or space based) supporting ADS-C (downlink of aircraft trajectory using EPP) as part of the ATS B2 services, achieving similar results, but has the downside of not having awareness of the FF-ICE/R2 message exchanges in the cockpit.

Linked to this trajectory downlink from the aircraft topic, there is a big difference between R&D activities on TBO happening in United States, Asia and Europe. While Europe is focusing on FF-ICE/R1 and Datalink to downlink trajectories ADS-C EPP, the other regions have been demonstrating the benefits of TBO based in FF-ICE/R2 and EFB systems to downlink the trajectory using IP connectivity instead of Datalink.

Both solutions for trajectory downlink from avionics are compatible, as ADS-C EPP can be used in aircraft forward-fitting while EFB fits better retro-fitting.

## III. REGULATORY FRAMEWORK

According to ICAO, the systems processing flight plans according to the FPL 2012 format should be sunset when sufficient experience with FF-ICE/R1 is gained and all necessary tools for deployment of full FF-ICE/R1 are in place. This is expected to happen early 2030's.

At the time of this writing, TBO-related regulations are in place only in the European region.

1. COMMISSION IR (EU) 2021/116 on the initial trajectory information sharing based on the aircraft downlinking trajectory information directly from the FMS to the ground systems via ATS-B2 ADS-C EPP. Deadline December 31st, 2027.
2. PCP Regulation No. 716/2014 & COMMISSION IR (EU) 2021/116 on Airspace Users and ANSPs to support FF-ICE/R1 filing services. Deadline December 31st, 2025.

## IV. EXPERIMENTAL ACTIVITIES IN TBO AND FF-ICE

TBO, either in part through early implementations, or in full Baseline (B) 2 deployment, will unlock efficiency gains through:

1. Resolved network capacity issues;
2. Improved metering to runway;
3. Improved flight path efficiency (Horizontal efficiency & Vertical/wind optimal efficiency);
4. Increased capacity that reduces delay and block time;

5. Reduced (enroute) delays and block time through increased predictability;
6. Improved ability to react to convective weather events; and
7. Increased ability to add flights.

Boeing recognizes these advances and fully supports their timely implementation, supporting efforts to mitigate operational and technical risks associated with TBO [6]. It is of foremost importance to demonstrate the feasibility and value of early TBO implementation. The following sections show examples of such activities.

*A. Flight test: Boeing ecoDemonstrator Explorer*

Over the past decade, Boeing's ecoDemonstrator program has been dedicated to accelerating innovation. It achieves this by taking promising technologies out of the lab and testing them in an operational environment. The program aims to solve real-world challenges faced by airlines, passengers, and the environment. Since its launch in 2012, this program has utilized eleven aircraft as flying test beds [8].

In 2023, Boeing collaborated with the Federal Aviation Administration (FAA), AEROTHAI, Civil Aviation Authority of Singapore (CAAS), and Japan Civil Aviation Bureau (JCAB) for a multiregional TBO project. The objective was to showcase and collaborate on TBO concepts [9].

The ecoDemonstrator Explorer 787 Dreamliner flight successfully passed through 10 airports, 5eASPs, and 4eAUs, highlighting the possibility of early TBO implementation using existing equipment and unlocking significant benefits, being the first oceanic flight ever supported by FF-ICE/R2 tools on the ground and in the cockpit.

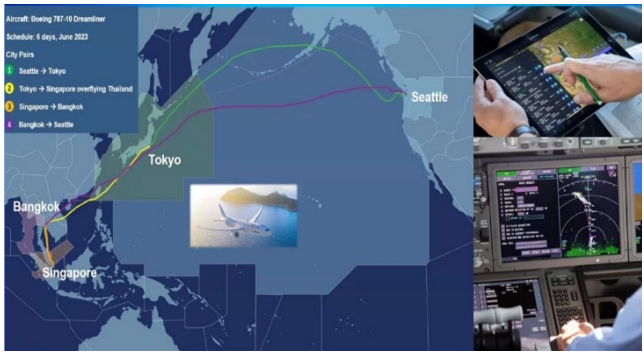

Figure 2: Boeing ecoDemonstrator Explorer - TBO Flights on June 2023

This activity demonstrated the utilization of ground systems infrastructure, modern aircraft with existing equipment, IP connectivity, and portable EFB to facilitate the sharing, negotiation, management, and utilization of trajectories with multiple Air Navigation Service Providers. This enabled the achievement of an optimal flight path across multiple regions, considering factors such as weather conditions, air traffic, and airspace closures.

*B. Europe: FF-ICE R1 Implementation*

The European SWIM (NM B2B Services) has already operationally available FF-ICE/R1 services.

Promising R&D activities have been focused on adapting state of the art flight planning tools to produce fully compliant trajectories that include additional resolution in Climb/Descent profiles, speeds, weather used for optimization, aircraft mass and its evolution, estimated times over (ETOs). That is full trajectory information never shared before by the airline enabling support from the European Network Manager and ATC units involved.

Europe offers an excellent framework for Initial TBO impact evaluation as maintains in parallel FLP2012 and FF-ICE platforms, so FPL 2012 can be filed Operationally, FF-ICE/R1 in a Pre-Operational environment and for the same set of flights, the impact on the whole network can be evaluated.

## V. THE TBO-ENABLED FOC

The introduction of FF-ICE/R1 is not expected to bring about changes in the roles and responsibilities of the airspace user in the FOC. However, it will have an influence on the tools and procedures used in performing their duties. As the initial phase of FF-ICE implementation primarily deals with pre-departure activities, FOC tools will primarily focus on flight preparation, flight dispatch, and flight planning tasks in a similar way to current ICAO FPL 2012 FOC tools.

When moving to FF-ICE/R2, the Airline is much more involved in Operations, flight and fleet management are enhanced. Airlines can initiate rerouting trajectory negotiation on disruption at any phase of flight, optimizing operations.

The TBO-enabled FOC should include different trajectory optimization engines empowered with SWIM connectivity and FF-ICE/R1+R2 implementation.

Given the amount of information exchanged in FF-ICE/R2 environments, and thanks to the real-time constraint sharing between stakeholders (enabled by AIXM/WXXM/FLXM information formats) automation needs to be included into these tools in the form of automatic trajectory negotiation when a new constraint shows up or an existing one changes or is no longer valid. The TBO FOC also should be able to perform multiple trial requests in order to check the validity of trajectories with different optimization objectives.

High automation combined with clear situation awareness is the only way of not increasing FOC operator's workload when moving to a full TBO implementation. Although automation plays a big role, manual modification of trajectories is also available for the operator.

*A. Trajectory efficiency*

One of the functions of a FOC is to generate and get approved flight plans that contain the most efficient trajectories possible. There is a multitude of constraints acting on a flight's trajectory, which usually can be classified in the two major categories of airspace constraints and aircraft performance constraints. While the latter has small and slow variations throughout an aircraft's lifetime, the former varies much more often and significantly. This is why there are Airline Flight Planning Systems (AFPS) which constantly consume aeronautical information and, from that, generate legally flyable flight plans, which are submitted to the ANSP for approval. The ANSP may reject the plan if one or more of the following conditions occur: it does not meet the airspace constraints; it conflicts with another plan; or it is incompatible with flow management measures. In case of rejection, the airline will have to revise the plan and re-submit, and this sequence can re-iterate.

Table 1: FOC tools in TBO

| FOC in current operations | FOC in TBO |
|---|---|
| **File and forget pre departure flight planning FPL 2012 – tactical adjustments later needed.** | Full trajectory negotiation in all phases of flight (PRE-departure & POST-departure) |
| **Very limited performance information sharing** | Climb/Descent profiles, speed, weather, weight, ETOs…full trajectory information never shared before by the airline |
| **Low situation awareness especially after filing** | Real time situation awareness enables the FOC to handle disruptions (by rerouting, delay etc). TBO tools include more automation and optimization. |
| **Limited connectivity with the aircraft** | Thanks to the EFB, the FOC has the capability to chat with the cockpit, receive EPP-Like (EFB ADS) status and trajectory message and negotiating trajectories. |

This section focuses on trajectory efficiency, in which efficiency consists of minimizing the Direct Operating Cost (DOC) of a flight. The DOC may include cost of fuel, cost of time, and Air Navigation fees, however, for simplicity, we will focus our analysis on fuel.

Minimizing a flight trajectory's DOC can be mathematically as tough as one wishes, but there are several simplifications used in practice that lead to the current day's levels of fuel efficiency, whose cost gap is believed to be in the ballpark of 8% more than which could be achieved with perfect use of all information available in the existing systems, concerning the aircraft and its operating environment. To fully close this gap, several systems must be changed, both onboard and in the ground, so that the technological cost is considerable. However, a smaller portion of this gap still may be closed with add-on optimizers which, by checking their solutions with service-proven systems, help to reap efficiency gains. This is explained by the case study below.

In this case study, the FOC has to generate a flight plan between Frankfurt (EDDF) and São Paulo / Guarulhos (SBGR). We can generate this flight plan with any AFPS, however it is important that it take into account all the existing airspace constraints. For flights over Europe, it has to respect the European Route Availability Documents (RADs) [10]. Most of the AFPS for real airline operations use proprietary algorithms which achieve some degree of optimization, however, it is widely known that one of the most efficient generic algorithms for solving the trajectory optimization problem is the so-called A* [11]. This algorithm relies on having a finite graph of waypoints, and is guaranteed to find the globally optimum path in time $O(b^d)$, where $b$ is the average branching factor, and $d$ the depth of solution. $b$ is determined by the airspace model, which is pretty much the same for every aircraft operator, and by the aircraft performance model, which determines which altitudes and maneuvers the aircraft can do; and $d$ limited by a certain bound, determined by the same airspace model and locations of origin and destination, roughly proportional to the distance between the two points. Thus, the secret sauce of such algorithms' performance is the heuristics used to discard branches in the search. But equally important is the quality of the models and of the input data, especially the weather forecast information, which determines the edge costs in the search graph.

Considering that the aircraft performance model and the weather model always need simplifications, especially regarding discretization and interpolation of multi-variable functions, we add a second optimization algorithm for the aircraft vertical profile, which explores a larger variety of possible trajectories. Then, we present different solutions for the pilots, so they can choose which one they are more comfortable to execute. For example, in a flight between the airports EDDF (Frankfurt) and SBGR (São Paulo / Guarulhos), the first stage of optimization resulted in the georeferenced path shown in Figure *3*.

That path is part of a 4-Dimensional (4-D) flight plan obtained from our AFPS. This is validated with the FF-ICE interface from NM B2B (Pre-Ops v. 27.0.0). The AFPS calculations indicated a total fuel of 56,655 kg. We consider only fuel amount due to the variable nature of fuel cost and cost of time among the different locations and airlines. We then ran the second optimization round, for the vertical profile. This second optimization algorithm, indicates a baseline fuel amount of 57,095 kg, and computed a new vertical profile which, if fully executed, could save around 3.6% of the cruise portion of the flight (~1,870 kg or 3.27% of the whole flight, reducing the total to 55,225 kg). However, this solution must be validated against airspace rules. This is done by means of the afore mentioned 1[st]-stage AFPS, by imposing certain input constraints to it. When so doing it, several errors are generated, because the plan is not compliant with the airspace rules. The input constraints are iteratively relaxed until a feasible plan is obtained, and this is pre-validated by the FPL 2012 interface. This plan is then converted to FF-ICE and validated by means of the NM FF-ICE trial request interface. The validated plan thus is different from the two previous stage plans, nevertheless with a small fuel saving in relation to the 1[st]-stage plan. These flight profiles are illustrated in Figure 4.

The fuel saving achieved by the final flight plan, in relation to the 1[st]-stage flight plan, is estimated as 169 kg of fuel, which is much smaller than what could be achieved with the ideal vertical profile, but still a significant figure, if this is done systematically for thousands of flights. The advantage provided by the FF-ICE validation, in relation to the legacy solution, which is based on FPL 2012, is that it allows to specify exactly where vertical change points occur, such as Bottom-Of-Climb (BOC), Top-Of-Climb (TOC), Top-Of-Descent (TOD) and Bottom-Of-Descent (BOD), which can be overlooked by the FPL 2012 interface, which looks at the vertical trajectory points as if they were disconnected from one another. This gives more assurance that the executed flight plan will be closer to the filed flight plan. In our example above, the airline dispatcher or pilot may choose between the two FF-ICE valid solutions with equal level of assurance.

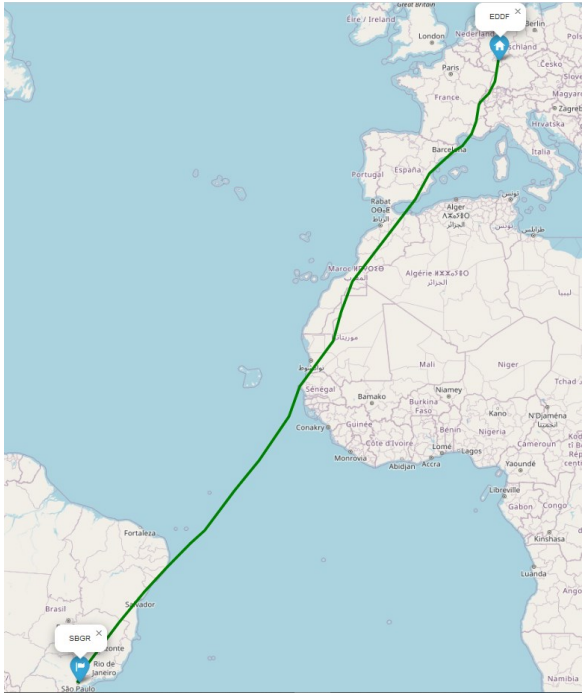

Figure 3: Horizontal path of the optimized route between EDDF (Frankfurt) and SBGR (São Paulo / Guarulhos) (Source: the authors).

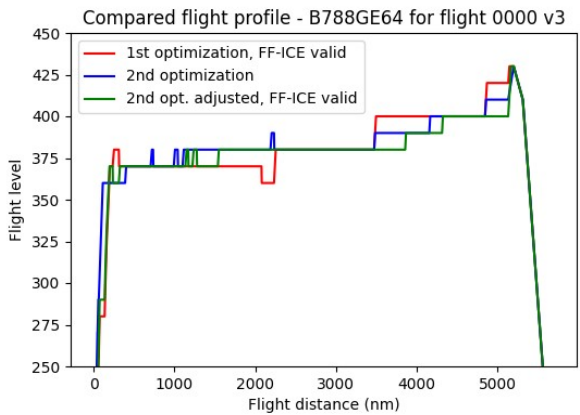

Figure 4: Comparison of the vertical profiles corresponding to the different optimization stages

*B. Automation (& Operational Efficiency)*

As mentioned in the previous section, FF-ICE flight plan validation helps to decrease the differences between filed and flown flight plans, by using a more accurate representation of the trajectory. This section provides a little more detail on this validation process.

In a case study, a hypothetical sample of 29 flights within the Eurocontrol NM jurisdiction was generated by our RAD-aware FPS. These flights were converted to FF-ICE trial requests and submitted to the NM FF-ICE Pre-Ops. V. 27.0.0 interface. The FF-ICE replies were collected and analyzed. The statistic in Figure 5 was then obtained.

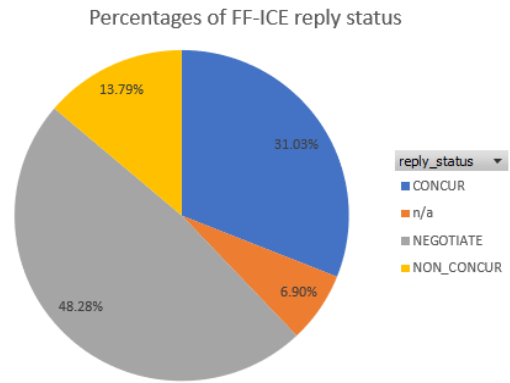

Figure 5: FF-ICE reply status from a sample of European flights.

The status CONCUR means that the flight plan is accepted as is, which was 31.03%, not a bad score. The status NON_CONCUR, which was only 13.79%, means that the plans have hard validity issues, which must be fixed before filing. The NEGOTIATE status, which represented 48.28% of the plans, means that, although the plan is legally valid and can be filed as is, NM is proposing a modification to better accommodate the flight according to discretionary rules, which are in place for ensuring higher airspace capacity, fluidity and other criteria stemming from Air Traffic Management (ATM) performance indicators. If the airline files such plans without changes, it is highly likely that, in later stages of the flight lifecycle, it will be requested to change by the ATM agent. These are rules not easily represented in the FPS, and having this instant feedback tremendously helps the airline dispatchers to make more assertive adjustments. The post-FF-ICE adjustment process can be greatly automated; however, we still because the FF-ICE responses have a well-defined data format, and contain explanatory messages which can be automatically processed.

The NON_CONCUR cases above are due to incompatibilities between the trajectory produced by the FPS and the FF-ICE validity rules, corresponding to 4 cases. Two of these cases generated messages which are actionable, in the sense that they indicate clearly what is wrong and can help either the dispatcher to change the flight plan manually in the FPS inputs or guide a developer to refine the FPS rules. Two other cases have messages which are difficult to interpret and thus not actionable at first encounter, with solution depending on a deeper investigation on the NM FF-ICE validation engine rules.

The "n/a" cases of Figure 5 happen when the incompatibility between the AFPS trajectory and the FF-ICE validity rules are caught during the generation of the FF-ICE trial request. More specifically, NM FF-ICE validation requires that vertical change points (VCP) be indicated at published and non-coincident waypoints, as for example, requiring that a new cruise level be indicated in a point distinct from the BOD and the TOC, and not before a previous TOC, which limits the possibilities for having consecutive vertical steps.

If, on the one hand, the AFPS should be capable of always finding flight plans compatible with FF-ICE validation rules, making the NON_CONCUR and conversion errors significantly rare; on the other hand it would be extremely costly to eliminate the NEGOTIATE cases, because such cases are due to transitory measures taken by the ATM agent, whose full processing would require large efforts for collecting real-time data, associated to highly complex algorithms for making optimization compatible with these transitory constraints. The unique value of having FF-ICE validation is to have instant and assertive feedback on the necessary changes to make the actual flight plan to adhere to the filed flight plan, feedback which can be used either by a human dispatcher, or an intelligent AFPS.

*C. Latency and performance*

In order to study the performance of FF-ICE messaging during the whole flight lifecycle, we switch our focus back to the ecoDemonstrator TBO flights, zooming in to the first leg of those flights, between Seattle and Tokyo. The performance aspects of this section are more related to data communication and the speed of negotiation of trajectory changes. We can break down the FF-ICE/R1+R2 message exchange by message type and ANSP during that flight. As it can be observed in Figure 6, message exchange happens during the whole flight and, beyond the negotiating ANSP that is selected in the FOC by applying the right negotiation horizon, agreed trajectory is distributed to all ANSP's in the path upfront the aircraft.

Trajectory negotiation can be started at the FOC but also the cockpit, as flight crew can do so from the EFB. That means that in terms of TBO performance in negotiation, multiple factors matter:

- Aircraft connectivity at any time during flight;
- EFB trajectory generation/evaluation from FF-ICE messages or the FMS;
- ANSP's SWIM connectivity;
- Controlling ANSP trajectory generation/evaluation performance;
- FOC trajectory generation/evaluation performance; and
- ANSP's SWIM connectivity.

As expected, when breaking down latencies by FF-ICE message type and ANSP/FOC times differ a lot depending on the complexity of operations to be performed after receiving the trajectory, anyway all lay down in an acceptable range of 0-3 seconds.

It is important to remark that analyzing the latency on individual messages do not make much sense as trajectory negotiation in most of the cases implies a conversation involving multiple messages.

In the case of Europe, the NM B2B services, latency for FF-ICE/R1 services available (TRIAL and FILING) is well beyond 1 second, not increasing with trajectory complexity in terms of number of information points.

One important remark on the latency and performance topic is aircraft connectivity. Aircraft is required to be connected the whole flight so the EFB can archive FF-ICE negotiation and FMS can execute Datalink clearances.

Although times during this demonstration were acceptable to implement initial TBO with minor modification on current technologies onboard, current initiatives regarding *Aircraft Multilink* and IPS connectivity are paving the future of aircraft communications to get a better performance and availability.

On top of the FF-ICE exchange, the EFB using ADI to interface avionics, has been downlinking trajectories to the FOC and relevant ANSPs to improve their predictions and provide constraints in the optimization process. Ground systems leverage FMS's ETO's, Top-of-climb, top-of-descent, actual and predicted weights and ETA to adjust trajectory predictions. That service, dubbed "EPP-like Data" (as EPP is Datalink ADS-C) had a higher latency [Standard Deviation: 7.5 sec], [Median Absolute Deviation: 2.0 sec], but its latency times do not appear to be correlated to any specific segment(s) of the flight. That latency is perfectly acceptable for FOC trajectory prediction engines to incorporate trajectory information from the EFB, extracted from avionics.

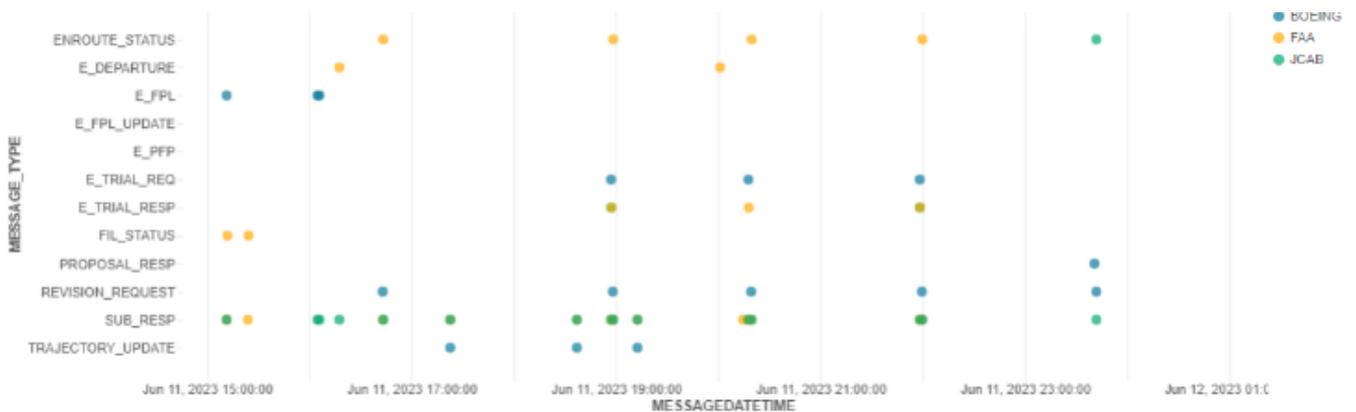

Figure 6: Boeing ecoDemonstrator Explorer 2023: Seattle to Tokyo flight FF-ICE message exchange

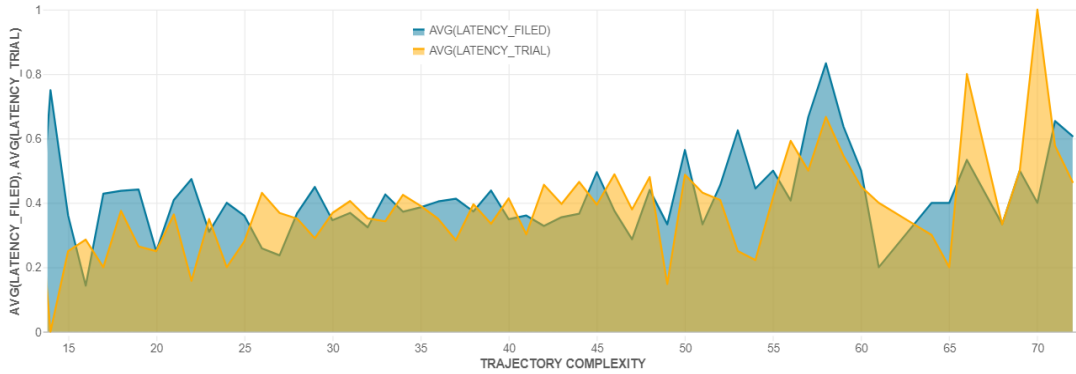

Figure 7: European NM B2B FF-ICE/R1 services latency.

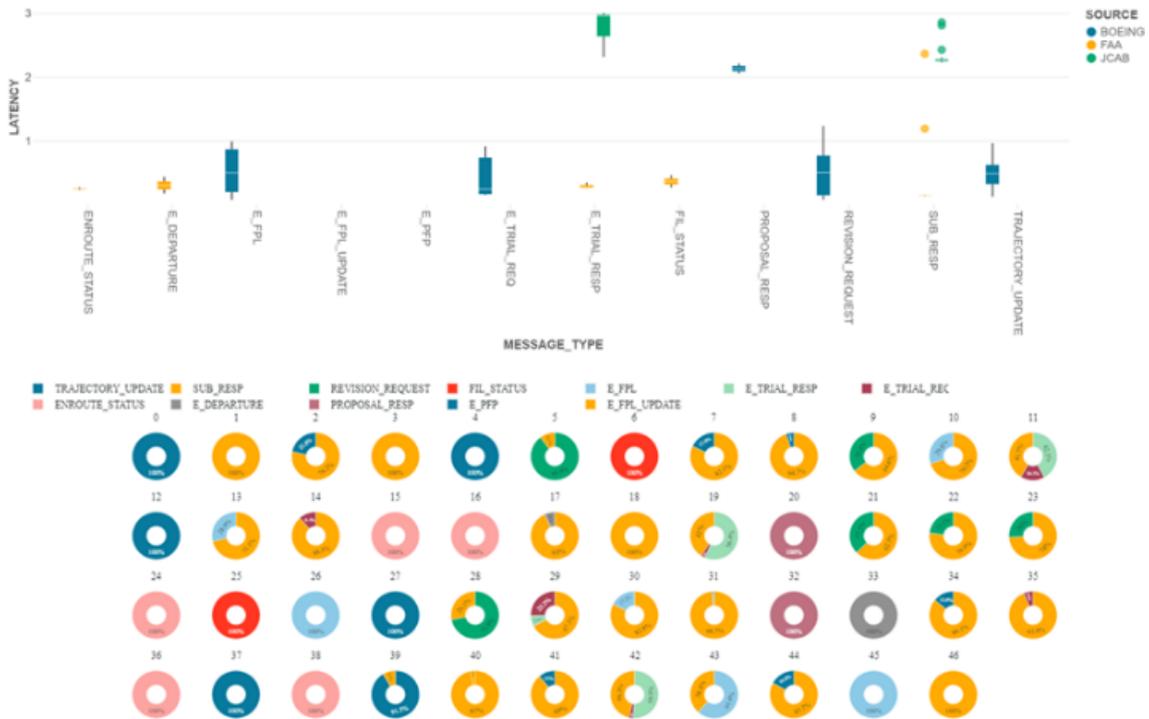

Figure 8: Latency times by FF-ICE message exchange on LEG1 ecoDemonstrator Explorer 2023

*D. Multi-ANSP trajectory negotiation*

During ecoDemonstrator Explorer 2023 activities, for FF-ICE/R2, the eAU provides the Revision Request to only the current controlling eASP unit. After the current controlling eASP unit Accepts the Revision Request, the eASP sends the Agreed Trajectory message to all downstream relevant eASPs. The Trajectory Update message communicates a small change to the flight plan that doesn't require a new clearance to downstream eASPs. The so-called Proposal Request is similar to a Trial Request but is initiated by the eASP. eASP to eASP trajectory negotiation is also in the scope of TBO but, as any other negotiation, the agreed trajectory must be shared with the FOC and new clearances sent to the cockpit.

## VI. Conclusion

In the authors' opinion, the study of the TBO-enabled FOC demonstrates the potential of sharing and coordinating trajectory information across multiple stakeholders, leading to enhanced efficiency. As demonstrated by the ecoDemonstrator Explorer 2023 test flights [5, 7], the accuracy of the trajectory optimization tools in the airline FOC was enhanced when consuming trajectory information downlinked by the EFB (obtained from avionics) and the latency times are acceptable, so initial TBO can happen with no big upgrades on the Airbourne or ground systems.

Also, automation plays a big role in FOC tools used at the TBO given the amount of information exchanged during all phases of flight.

The following Operational Values of TBO encompass the significant benefits delivered to both ASPs (Airspace Service Providers) and AUs (Aircraft Users):

**Enhanced Predictability**: Sharing a common plan among stakeholders improves predictability by operating based on the same trajectories. This reduces confusion and enhances overall operational efficiency.

**Increased, Reliable Flexibility**: The ability to accommodate trajectory changes while maintaining business objectives provides increased and reliable flexibility. For example, adjusting contingency fuel based on evolving conditions reduces fuel burn, emissions, and aircraft weight, while increasing payload and safety.

**Improved Strategic Planning**: Incorporating out-of-zone traffic and mitigating deterministic delay factors through improved planning enhances network performance. This approach also ensures more equitable distribution of delays.

**Decreased Uncertainty**: Improved trajectory accuracy reduces uncertainty within the system, leading to smoother operations and better decision-making.

**Alignment of Strategic Plans and Tactical Actions**: By utilizing the trajectory as a common framework, strategic planning and tactical actions can be aligned. This enables effective sharing, management, and utilization of trajectory information, resulting in improved overall performance.

These TBO Operational Values highlight the positive outcomes achieved during the demonstrations, and which will be expected when TBO gets widespread.

## VII. Acknowledgment

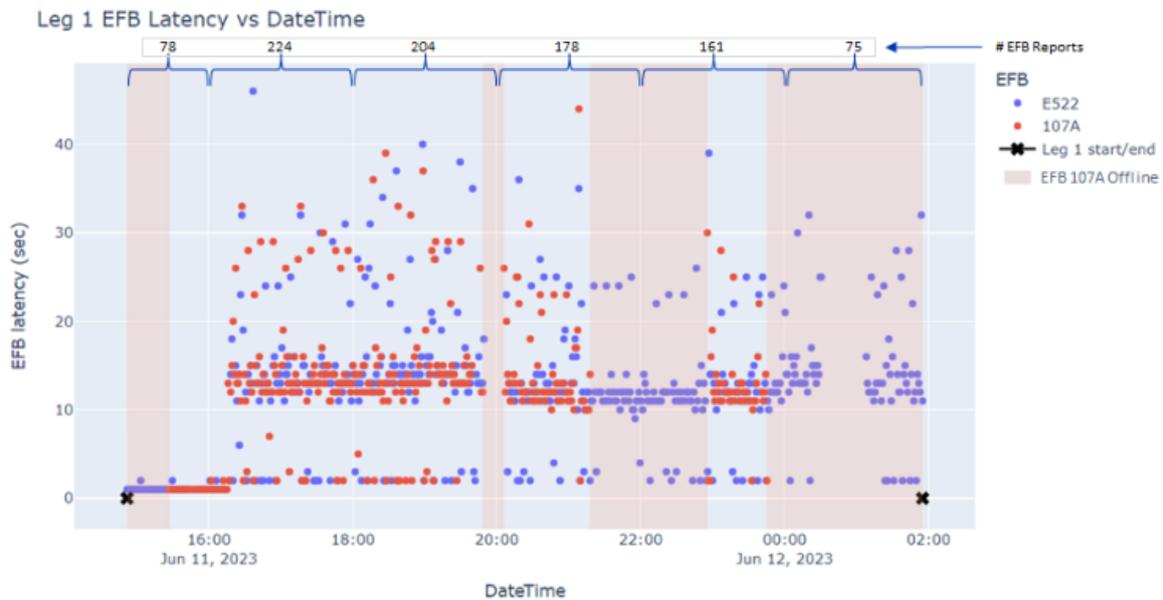

Figure 9: EFB latency on trajectory [EPP-like] downlink to the FOC